\documentclass[letterpaper, 10 pt,conference]{ieeeconf}  

\IEEEoverridecommandlockouts                              

\usepackage{graphics} 
\usepackage{epsfig} 
\usepackage{amsmath} 
\usepackage{amssymb}  
\usepackage{url,algorithm}
\usepackage{multirow}
\usepackage{color}
\usepackage{subfigure}
\usepackage{tikz}
\usepackage{pgfplots}

\let\labelindent\relax
\usepackage{enumitem}
\renewcommand{\theenumi}{\arabic{enumi}}

\renewcommand{\theenumii}{\arabic{enumii}}

\renewcommand{\theenumiii}{\arabic{enumiii}}

\renewcommand{\theenumiv}{\arabic{enumiv}}

\newcommand{\smallmat}[1]{\left[ \begin{smallmatrix}#1 \end{smallmatrix} \right]}

\newcommand{\argmin}{\mathop{\rm arg\ min}\nolimits}

\newtheorem{remark}{Remark}{\normalfont}{\normalfont}

\newcommand{\figurename}[1]{Figure{#1}}
\newcommand{\tablename}[1]{Table{#1}}

\title{\LARGE \bf Shrinkage Strategies for Structure Selection and Identification of Piecewise Affine Models   
}

\author{Valentina Breschi and Manas Mejari
	\thanks{This work has been accepted to the 59-th IEEE Conference on Decision and Control.}
		\thanks{V. Breschi is with  Dipartimento di Elettronica e Informazione, Politecnico di Milano, Milano, Italy. {\tt\small valentina.breschi@polimi.it}}
	\thanks{M. Mejari is with IDSIA Dalle Molle Institute for Artificial Intelligence, SUPSI-USI, Manno, Switzerland. {\tt\small manas.mejari@supsi.ch} }
}   %

\newcommand{\MM}[1]{\textcolor{black}{#1}}

\begin{document}

\maketitle
\thispagestyle{empty}
\pagestyle{empty}

\begin{abstract}
We propose two \emph{optimization-based} heuristics \MM{for structure selection and  identification of}  \emph{PieceWise Affine} (PWA) models with exogenous inputs. \MM{The first method determines the number of affine sub-models assuming known model order of the sub-models, while the second approach estimates the model order for a given number of affine sub-models.}
Both approaches rely on the use of \emph{regularization-based} shrinking strategies, that are exploited within a \emph{coordinate-descent} algorithm. This allows us to estimate 
\MM{the structure of the PWA models along with its model parameters.} 
Starting from an over-parameterized model, the key idea is to alternate between an identification step and structure refinement, based on the sparse estimates of the \MM{model} parameters. The performance of the presented strategies is assessed over two benchmark examples.  
\end{abstract}
\section{Introduction}
Many real-world systems and phenomena are characterized by different operating conditions, among which they can commute smoothly and/or abruptly. In such scenarios, linear models are usually insufficient to accurately describe the behavior of the underlying system and one must resort to more complex model structures to attain satisfactory approximations. However, many learning frameworks lead to models that are rather hard to analyze and use, especially for control purposes.

A good trade-off between complexity, approximation power and intelligibility can be attained by considering \emph{piecewise affine} (PWA) maps. Indeed, it is well known that PWA functions have universal approximation properties \cite{Lin1992,Breiman1993}, while featuring a rather simple structure, since they are characterized by a finite collection of affine sub-models defined over a polyhedral partition of the regressor space. Thus, PWA models can be analyzed and exploited for control design with state-of-the-art techniques (see \emph{e.g.,} \cite{BEMPORAD1999}).

Although appealing because of the aforementioned properties, PWA models are challenging to learn from data. Indeed, to identify the local models and estimate the associated regions of the regressor space one should know the operating condition of the real systems at each time step, which is usually not available a priori. The learning problem thus entails not only the identification of sub-models parameters and the computation of a polyhedral partition of the regressor space, but it also requires one to solve an unsupervised clustering problem, since each sample has to be assigned to one of the local models. 

Several heuristics have been proposed over the years to deal with this demanding learning task, among which we mention the bounded-error method proposed in \cite{Bemporad2005}, the clustering-based procedures in \cite{BAKO2011_2,BBPAut15,FERRARITRECATE2003,NAKADA2004}, the mixed-integer programming based approaches in \cite{Naik2017,ROLL2004}, the Bayesian method in \cite{Juloski2005} and the sparse optimization techniques proposed in \cite{BAKO2011,OHLSSON2013}. However, most of these methods rely on the prior knowledge on the number of operating conditions of the true system and of its order, which might not be available in practice. In \cite{Bemporad2005}, the number of sub-models is chosen by solving a \emph{minimum} number of \emph{feasible subsystems} (MIN PFS) problem, which is, in turn, formulated by relying on a bound for the estimation error. This parameter is unlikely to be available in practice, thus becoming a tuning knob for the algorithm. A refinement strategy is also proposed therein, where the number of sub-models is further reduced by merging local models that share the same parameters and discarding the ones that are associated to \textquotedblleft few\textquotedblright \ data-points. By exploiting sparse optimization, the methods presented in \cite{BAKO2011,OHLSSON2013} both allows for the selection of the number of modes. Specifically, the approach proposed in \cite{OHLSSON2013} considers an over-parameterized model and, then, exploits $L_{1}$-regularization to enforce the local parameter estimates to be similar. Instead, in \cite{BAKO2011}, $L_{1}$-regularization is used to iteratively search for the sparsest vector of fitting errors to cluster the data-point, estimate local models and decide on their number. To the best of our knowledge, only in \cite{BAKO2011}, it is provided a possible solution for the selection of the sub-models structure, which relies on the introduction of an additional $L_1$-regularization term on the parameters.

As an alternative, in this paper we present two \emph{regularization-based} strategies that allow the partial selection of the PWA model structure along with parameter estimation. These techniques exploit the flexibility of the general purpose approach for jump model fitting presented in \cite{BPBB2018}, which can easily incorporate regularization-based shrinking strategies. In particular, we propose a heuristic to estimate the number of local models by using $L_\infty$ regularization on the sub-models parameters, while accounting for the cardinality of the associated clusters to refine the optimization-driven choice. In the second approach, the local model structure is estimated via the \emph{elastic net}. We remark that the proposed strategy is inspired by the one briefly discussed in \cite{BAKO2011}, but it inherits the benefits of elastic net over simple lasso regularization. The overall identification procedure with partial structure selection requires the alternation between a learning stage and a step in which the model is refined. Once the model structure is determined, the final model is re-estimated by neglecting the shrinkage regularization terms. Although the proposed strategies are quite general and can be straightforwardly extended to other classes of PWA models, in this work we focus on \emph{piecewise affine autoregressive models with exogenous inputs} (PWARX).     

The paper is organized as follows. The problem of learning a PWARX model with partially unknown structure is introduced in Section~\ref{Sec:1} and the core algorithm used to address this learning task is summarized in Section~\ref{Sec:2}. The proposed regularization-based strategies are then discussed in Section~\ref{Sec:3}, and their effectiveness is shown in Section~\ref{Sec:4} on a numerical example and an experimental case study. Conclusions and directions for future work are finally discussed in Section~\ref{Sec:Conclusions}. 

\section{Problem formulation}\label{Sec:1}
Consider the following \emph{dynamical} data-generating system, whose instantaneous response $y_{t} \in \mathcal{Y} \subseteq \mathbb{R}$ to an input $u_{t} \in \mathcal{U} \subseteq \mathbb{R}$ is given by
\begin{subequations}
\begin{equation}\label{eq:data_generating}
y_{t}=f_{\mathrm{o}}(x_{t})+e_{t}^{\mathrm{o}},
\end{equation}
where $e_{t}^{\mathrm{o}} \in \mathbb{R}$ is a zero-mean additive white noise independent of the \emph{regressor} $x_{t} \in \mathcal{X} \subseteq \mathbb{R}^{n_x}$. Let the regressor be given by the following collection of past input/output samples
\begin{equation}\label{eq:regressor}
x_{t} =\begin{bmatrix}
y_{t-1} & \cdots & y_{t-n_{a}} & u_{t-1} & \ldots u_{t-n_{b}}
\end{bmatrix}',
\end{equation}  
\end{subequations}
with $n_{a},n_{b} \in \mathbb{R}^{+}$ denoting the \emph{possibly unknown} order of the system. The \emph{unknown} function $f_{\mathrm{o}}: \mathcal{X} \rightarrow \mathbb{R}$ characterizing the underlying regressor/output relationship is assumed to be \emph{nonlinear} and \emph{possibly discontinuous}, but not necessarily piecewise affine. 

\MM{We aim at finding a PWA approximation $f$ of the unknown function $f_{\mathrm{o}}$, based on a set of inputs $U=\{u_{t}\}_{t=1}^{T}$ and output samples  $Y=\{y_{t}\}_{t=1}^{T}$, generated by \eqref{eq:data_generating}-\eqref{eq:regressor}. We remark that, due to the universal approximation properties of PWA maps~\cite{Lin1992,Breiman1993}, PWA function $f$ can approximate $f_{\mathrm{o}}$ with an arbitrary accuracy.}

The approximating PWA function $f: \mathcal{X} \rightarrow \mathbb{R}$ \MM{is defined as}
\begin{equation}\label{eq:approx_map}
f(x_{t})=\begin{cases}
\tilde{x}_{t}'\theta_{y,1}, \quad \mbox{if } x_{t} \in \mathcal{X}_{1},\\
\vdots\\
\tilde{x}_{t}'\theta_{y,K}, \quad \mbox{if } x_{t} \in \mathcal{X}_{K},
\end{cases}
\end{equation} 
with $\tilde{x}_{t}=\smallmat{x_{t}' & 1}'$. This \emph{piecewise affine autoregressive} model with \emph{exogenous inputs} (PWARX) is a collection of $K \in \mathbb{N}$ local affine models, characterized by the parameters  $\Theta_{y}=(\theta_{y,1},\ldots,\theta_{y,K})$  and defined over a \emph{complete} polyhedral partition\footnote{$\{\mathcal{X}_{k}\}_{k=1}^{K}$ is a complete polyhedral partition of $\mathcal{X}$ if $\bigcup_{k=1}^{K} \mathcal{X}_{k}=\mathcal{X}$ and $\overset{\circ}{\mathcal{X}_{i}} \cap \overset{\circ}{\mathcal{X}_{j}}=\emptyset$, for $i \neq j$, $i,j=1,\ldots,K$, with $\overset{\circ}{\mathcal{X}_{i}}$ denoting the interior of the $i$-th polyhedron $\mathcal{X}_{i}$.} $\{\mathcal{X}_{k}\}_{k=1}^{K}$ of the regressor space $\mathcal{X}$. We stress that all the local models share the same structure by definition.

We characterize the partition through a piecewise affine separator function $\phi: \mathbb{R}^{n_x} \rightarrow \mathbb{R}$, defined as
	\begin{equation}\label{eq:pwa_seperator}
	\phi(x) = \max_{k=1,\ldots,K} \left\{\theta_{x,k}' \tilde{x}\right\}, 
	\end{equation}
	where $\theta_{x,k}' \in \mathbb{R}^{n_x+1}$. Thus, a polyhedron $\mathcal{X}_{i}$ can be described by the following linear inequalities
	\begin{equation}\label{eq:polyhedron}
	\mathcal{X}_{i}\!=\!\{x \in \mathbb{R}^{n_x}: (\theta_{x,i}-\theta_{x,j})' \tilde{x} \geq 0, j=1,\ldots,K,j \neq i\}.
	\end{equation}

Therefore, learning a PWARX model from data amounts at: $(i)$ choosing the number of affine sub-models $K$; $(ii)$ estimating the parameters $\Theta_{y}$ of the local models, $(iii)$ finding the parameters $\Theta_{x}=(\theta_{x,1},\ldots,\theta_{x,K})$ characterizing the PWA separator in \eqref{eq:pwa_seperator}, and, eventually, $(iv)$ selecting the order of the sub-models, namely $n_{a}$ and $n_{b}$, when it is unknown. 

In this work, we split model structure selection into two separate tasks by assuming that \emph{either} the number of local models $K$ \emph{or} the sub-model order is fixed a priori. In the first task, we estimate $(\Theta_{y},\Theta_{x},K)$ for fixed $n_{a},n_{b}$, while in the second one we fix $K$ and retrieve $(\Theta_{y},\Theta_{x},n_{a},n_{b})$ from data.

\section{Learning PWARX model}\label{Sec:2}
It is well known that learning PWA approximating map is quite challenging, since this problem is NP-hard~\cite{LAUER2015148} even for an a-priori fixed model structure. Indeed, it requires the concurrent solution of local linear regressions, a classification and a clustering problem. To cope with this, we recall the \emph{modular} optimization problem presented in \cite[Section 2.2.1.]{BPBB2018} for PWA regression and we summarize the steps of the \emph{coordinate-descent} approach proposed therein to solve the problem at hand. 

Let $X=\{x_{t}\}_{t=1}^{T}$ be the sequence of regressors constructed from the available input/output dataset $\{U,Y\}$. Let $s_{t} \in \{1,\ldots,K\}$ be the \emph{latent} variable, that indicates which local model is associated with the $t$-th regressor/output pair $\{\tilde{x}_{t},y_{t}\}$, such that the sequence $\mathcal{S}=\{s_{t}\}_{t=1}^{T}$ comprises the label associated to each data point. 

Within the optimization-based framework of \cite{BPBB2018}, learning a generic \emph{jump} model entails the solution of the following optimization problem:
\begin{subequations}
\begin{equation}\label{eq:general_problem}
\begin{aligned}
\min_{\Theta_{y},\Theta_{x},\mathcal{S}}~~ \sum_{t=1}^{T} &\left[ \ell_{y}(y_{t},x_{t},\theta_{y,s_t})+\rho \ell_{x}(x_{t},\theta_{x,s_t})\right]+\\
&\quad\quad\quad+\sum_{k=1}^{K} \left[r_{y}(\theta_{y,k})+\rho r_{x}(\theta_{x,k})\right], 
\end{aligned}
\end{equation} 
In particular, for the PWARX model class \eqref{eq:approx_map}-\eqref{eq:pwa_seperator}, the \emph{loss function} $\ell_{y}: \mathcal{X} \times \mathcal{Y} \times \mathbb{R}^{n_x+1} \rightarrow \mathbb{R} \cup \{+\infty\}$ is chosen as the squared \emph{local fitting error}, \emph{i.e.,}
\begin{equation}
\ell_{y}(y_{t},x_{t},\theta_{y,s_t})=(y_{t}-\tilde{x}_{t}'\theta_{y,s_t})^{2},
\end{equation}
and the loss function $\ell_{x}: \mathcal{X} \times \mathbb{R}^{n_x+1} \rightarrow \mathbb{R} \cup \{+\infty\}$ is
\begin{equation}
\ell_{x}(x_{t},\theta_{x,s_t})=\sum_{\substack{j=1\\j \neq s_{t}}}^{K} \max{\left\{0,(\theta_{x,j}\!-\!\theta_{x,s_t})'\tilde{x}_{t}\!+\!1\right\}}^{2},
\end{equation}
so to account for the violations of the inequalities in \eqref{eq:polyhedron}.
The \emph{regularizers} $r_{y}: \mathbb{R}^{n_x+1} \rightarrow \mathbb{R} \cup \{+\infty\}$ and $r_{x}: \mathbb{R}^{n_{x}+1} \rightarrow \mathbb{R} \cup \{+\infty\}$ act on the parameters $\Theta_{y}$ and $\Theta_{x}$, characterizing the local models and the PWA separator respectively. The positive hyper-parameter $\rho$ balances the effect of each term on the overall cost.
\end{subequations}
From \eqref{eq:general_problem}, it is clear that the discrete sequence $\mathcal{S}$ couples the regularized linear regressions and the multi-category discrimination problems, that have to be solved to find $\Theta_{y}$ and $\Theta_{x}$, respectively. Nonetheless, once $\mathcal{S}$ is fixed, the parameters $\Theta_{y}$ and $\Theta_{x}$ can be found independently from one another. On the other hand, for a given $\Theta_{y}$ and $\Theta_{x}$, the discrete sequence $\mathcal{S}$ can be computed using \emph{dynamic programming} (DP) \cite{bertsekas1999nonlinear} technique in a computationally efficient manner. 

Based on the above considerations, we exploit the \emph{coordinate-descent} approach proposed in \cite{BPBB2018} by alternatively solving problem~\eqref{eq:general_problem} with respect to $\Theta_{y}$ and $\Theta_{x}$, for a fixed sequence $\mathcal{S}$, and optimizing with respect to $\mathcal{S}$ for fixed $\Theta_{y}$ and $\Theta_{x}$. The three step procedure tailored for the solution of problem \eqref{eq:general_problem} is summarized in Algorithm~\ref{algo1} where, with a slight abuse of notation, we denote
\begin{subequations}
\begin{align}
&\ell_{y}(Y,X,\Theta_{y},\mathcal{S})=\sum_{t=1}^{T} \ell_{y}(y_{t},x_{t},\theta_{y,s_t}),\\
&\ell_{x}(X,\Theta_{x},\mathcal{S})=\sum_{t=1}^{T} \ell_{x}(x_{t},\theta_{x,s_t}),\\
&r_{y}(\Theta_{y})=\sum_{k=1}^{K} r_{y}(\theta_{y,k}), \quad r_{x}(\Theta_{x})=\sum_{k=1}^{K} r_{x}(\theta_{x,k}).
\end{align}
\end{subequations}
Given the structure of the local models, the problem ought to be solved at step~\ref{step:loc_mod} can be split into $K$ distinct regularized linear regression problems, one for each local models. We remark that each sub-model is updated by using only the data points that are associated to it, according to the estimated mode sequence $\mathcal{S}^{i-1}$. The estimation of the PWA separator at step~\ref{step:part} entails the solution of a multi-category discrimination problem, which can be efficiently computed with the Newton-like method proposed in \cite{BBPAut15}. The mode sequence is finally updated at step~\ref{step:seq} via DP \cite{bertsekas1999nonlinear}, as detailed in \cite{BPBB2018}, by neglecting the regularization terms as they are independent of $\mathcal{S}$. Note that, the tunable parameter $\rho \in \mathbb{R}^{+}$ is not considered when optimizing with respect to $\Theta_{y}$ and $\Theta_{x}$ (steps~\ref{step:loc_mod}-\ref{step:part}), but it is exploited when updating the mode sequence to trade-off between the fitting error and the cost of choosing a certain polyhedral region.
\begin{algorithm}[!tb]
	\caption{[Coordinated descent for learning PWARX \cite{BPBB2018}]}
	\label{algo1}
	~~\textbf{Input}: Data $\{Y,X\}$; initial mode sequence $\mathcal{S}^{0}$; $\rho \in \mathbb{R}^{+}$.
	\vspace*{.1cm}\hrule\vspace*{.1cm}
	\begin{enumerate}[label=\arabic*., ref=\theenumi{}]
		\item \textbf{for} $i=1,\ldots$ \textbf{do} \vspace{.1cm}
		\begin{enumerate}[label=\theenumi{}.\arabic*., ref=\theenumi{}.\theenumii{}]	
			\item[$\triangleright$] Estimate local functions 	\vspace{.2cm}			%
			\item \label{step:loc_mod}   $\Theta_{y}^{i} \leftarrow \argmin_{\Theta_{y}}  \ell_{y}(Y,X,\Theta_{y},\mathcal{S}^{i-1})+r_{y}(\Theta_{y})$ \vspace{0cm}
			\item[$\triangleright$] Estimate linear separators\vspace{.2cm}
			\item \label{step:part} $\Theta_{x}^{i} \!\leftarrow\! \argmin_{{\Theta}_{x}} \ell_{x}(X,\Theta_{x},\mathcal{S}^{i-1})+r_{x}(\Theta_{x})$\vspace{.2cm}
			
			\item[$\triangleright$] Estimate mode sequence\vspace{.2cm}
			\item \label{step:seq} $\mathcal{S}^{i} \!\leftarrow\! \argmin_{\mathcal{S}} \ell_{y}\left(Y,X,\Theta_{y}^{i},\mathcal{S}\right)+\rho \ell_{x}(X,\Theta_{x}^{i},\mathcal{S})$\vspace{.2cm}
		\end{enumerate}
		\item \label{step:stop crit} \textbf{until} $\mathcal{S}^{i}=\mathcal{S}^{i-1}$
	\end{enumerate}
	\vspace*{.1cm}\hrule\vspace*{.1cm}
	~~\textbf{Output}: Local models $\Theta_{y}^{\star}\!=\!\Theta_{y}^{i}$; PWA separator $\Theta_{x}^{\star}\!=\!\Theta_{x}^{i}$; sequence $\mathcal{S}^{\star}=\mathcal{S}^{i}$.
\end{algorithm}
\begin{remark}[Inference hints]
	Once a PWARX model is retrieved by running Algorithm~\ref{algo1}, for each new regressor $x_{t}$ the active local model can be computed as
	\begin{subequations}
	\begin{equation}
	\hat{s}_{t}=\max_{k=1,\ldots,K} \left\{(\theta_{x,k}^{\star})'\tilde{x}_t \right\},
	\end{equation} 
	and, accordingly, the corresponding output can be obtained as follows:
	\begin{equation}
	\hat{y}_{t}=\tilde{x}_{t}'\theta_{y,\hat{s}_{t}}^{\star}.
	\end{equation}
	\end{subequations} 
	\hfill $\blacksquare$
\end{remark}

\section{Regularization-based strategies for model structure selection}\label{Sec:3}
Suppose that either the number of sub-models characterizing the PWA map in \eqref{eq:approx_map} or their local structure is \emph{unknown} and assume that bounds on the values of these unknowns are given by the maximum tolerable complexity of the estimated model.

To determine the unknown PWARX model structure (\emph{i.e.,} either the number of sub-models $K$ or their model order $n_{a}$, $n_{b}$), the regularization terms in the objective function of \eqref{eq:general_problem} can be shaped so to use well-known \emph{regularization-based} shrinkage techniques to choose the structure of the estimated PWARX model in \eqref{eq:approx_map}, directly from data. This allows to avoid grid searches, which usually require an exhaustive exploration of the unknown parameter space.

In this work, starting from an over-parameterized model, we aim at devising iterative structure selection strategies that rely on runs of Algorithm~\ref{algo1} with tailored regularization terms. In particular, we focus on the selection of the \emph{group regularization} $r_{y}(\Theta_{y})$, which acts on the parameters of the local models.

 \MM{Two different choices of $r_{y}(\Theta_{y})$ are made to tackle the problems of estimating the number of sub-models $K$ and the model order $n_{a}, n_{b}$ respectively. For both problems, we fix the regularizer $r_{x}(\Theta_{x})$ acting on the parameters $\Theta_{x}$ of the PWA seperator as follows}
\begin{equation}\label{eq:PWA_reg}
r_{x}(\Theta_{x})=\lambda \sum_{k=1}^{K} \|\theta_{x,k}\|_{2}^{2},
\end{equation}
with $\lambda \in \mathbb{R}^{+}$ being a hyper-parameter to be tuned. This choice makes the multi-category discrimination optimization at step~\ref{step:part} of Algorithm~\ref{algo1} strictly convex, thus reducing the complexity of the resulting model. In the following, we introduce two different strategies based on the choice of $r_{y}(\Theta_{y})$, respectively tailored to handle the selection of $K$, for fixed local model structure, and the choice of $n_{a}$ and $n_{b}$, for a given number of local models. In the latter scenario, we assume that $n_{a}, n_{b} \geq 1$, namely that the response of the system at a given instant always depends at least on the previous input/output pair. In both cases, we assume that the dataset is relatively \emph{balanced}, namely we expect that the local models are associated to a reasonable number of samples for the overall model to be identifiable with sufficient accuracy.

\subsection{Number of local model selection via $L_{\infty}$ regularization}
\MM{In this subsection, we outline the strategy to estimate number of local models $K$ for a fixed model orders $n_{a}, n_{b}$. Let us assume that at most $K_{max}$ local models in \eqref{eq:approx_map} are sufficient to accurately describe $f_{\mathrm{o}}$ in \eqref{eq:data_generating} and $K_{max}$ is known}. As a first step towards the definition of a strategy to select the number of sub-models, we introduce a regularization term $r_{y}(\Theta_{y})$ that shrinks the entire parameter vectors $\theta_{y,k}$ towards zero, when the corresponding sub-models are redundant. To this end, we choose $r_{y}(\Theta_{y})$ to be a convex combination of \emph{group Tikhonov} regularization and a term penalizing the mixed $L_{1-\infty}$-norm of the parameter vectors, \emph{i.e.,} 
\begin{equation}\label{eq:l_infty_2}
r_{y}(\Theta_{y})=\mu \sum_{k=1}^{K} \|\theta_{y,k}\|_{2}^{2}+\nu \sum_{k=1}^{K} \|\theta_{y,k}\|_{\infty},
\end{equation}
where $\mu, \nu \in \mathbb{R} $ are hyper-parameters to be tuned. The term penalizing the sum of the infinity norms (\emph{i.e.,} the mixed $L_{1-\infty}$-norm), ensures that, at the solution, the vector $\theta_{y,k}$ is either full or have all elements nearly identical to zero. Thus, only the component of $\theta_{y,k}$ with the highest absolute value affects the cost. The additional $L_2$ regularization term is introduced for a better numerical conditioning of the problem when the parameter vector is not shrunk towards zero.

Since the shrinkage strategy might lead to parameter vectors with rather small elements, that are still not identically zero, we label a local model as \emph{redundant} if the associated parameter vector $\theta_{y,k}$ satisfies the following inequality: 
\begin{equation}\label{eq:check1}
\|\theta_{y,k}\|_{\infty}\leq \delta, 
\end{equation}  
where $\delta \in \mathbb{R}^{+}$ is a relatively small hyper-parameter to be tuned. 

Accordingly, starting from an over-parameterized model with $K=K_{max}$, the procedure for the automatic selection of the model structure is summarized in Algorithm~\ref{algo2}. At the beginning of each iteration, Algorithm~\ref{algo1} is run with the regularization term in \eqref{eq:l_infty_2}. Therefore, the estimation of the local models for a fixed mode sequence $\mathcal{S}$ (step~\ref{step:loc_mod} of Algorithm~\ref{algo1}) entails the solution of the following problem:
 \begin{equation}
 \min_{\Theta_{y}} \sum_{t=1}^{T} \sum_{k=1}^{K} (y_{t}\!-\!\tilde{x}_{t}'\theta_{y,s_t})^{2}\!+\!\mu \sum_{k=1}^{K} \|\theta_{y,k}\|_{2}^{2}\!+\!\nu \sum_{k=1}^{K} \|\theta_{y,k}\|_{\infty}.
 \end{equation}  
Once the new PWARX model is retrieved (\emph{i.e.,} once algorithm~\ref{algo1} has stopped), redundant sub-models are detected by checking the condition in \eqref{eq:check1} for the estimated parameters $\Theta_{y}^{\star}$ and the number of local models is updated accordingly (see steps~\ref{step:start-red} and \ref{step:red}). Specifically, at step ~\ref{step:start-red}, the set of indexes $\mathcal{R}^{j}$ corresponding to the redundant sub-models is updated based on the criterion \eqref{eq:check1} and at \ref{step:red}, the number of local models $K^j$ is updated by subtracting the number of redundant sub-models (\emph{i.e.} the cardinality $\#\mathcal{R}^{j}$ of set  $\mathcal{R}^{j}$) from the number of sub-models estimated $K^{j-1}$ at the previous iteration. If no redundant model is detected (\MM{$\#\mathcal{R}^{j} =0$}), at steps~\ref{step:clustbuilt}-\ref{step:modup} we further check the clusters $\{\mathcal{C}_{k}\}_{k=1}^{K}$ induced by $\mathcal{S}$, \emph{i.e.,} 
\begin{equation}
\mathcal{C}_{k}=\{x_{t} \in X: s_{t}^{\star}=k\}, \quad k=1,\ldots,K
\end{equation}
and we exploit the assumption that the dataset is balanced to remove as many local models as the number of \emph{almost empty} clusters, namely sets with cardinality $\#\mathcal{C}_{k}$ smaller or equal than $1\%$ of the overall dataset. This procedure is iterated until $K$ does not vary over two consecutive iterations.    

Once Algorithm~\ref{algo2} is terminated a new instance of Algorithm~\ref{algo1}  is run with $r_{y}(\Theta_{y})$ defined as in \eqref{eq:l_infty_2} with $\nu=0$ to identify the final PWARX model with number of local models $K^{\star}$.   

\begin{algorithm}[!tb]
	\caption{[Selection of the number of sub-models $K$]}
	\label{algo2}
	~~\textbf{Input}: Dataset $\{Y,X\}$ of length $T$; initial mode sequence $\mathcal{S}^{0}$; $\rho, \mu, \nu, \lambda, \delta \in \mathbb{R}^{+}$; maximum number of local models $K^{0}=K_{max}$.
	\vspace*{.1cm}\hrule\vspace*{.1cm}
	\begin{enumerate}[label=\arabic*., ref=\theenumi{}]
		\item \textbf{for} $j=1,\ldots$ \textbf{do} \vspace{.1cm}
		\begin{enumerate}[label=\theenumi{}.\arabic*., ref=\theenumi{}.\theenumii{}]	
			\item \label{step:algo1K} \textbf{run} Algorithm~\ref{algo1} for $K\!=\!K^{j\!-\!1}$ and $r_{y}(\Theta_{y})$ as in \eqref{eq:l_infty_2}; 
			\item \label{step:start-red}$\mathcal{R}^{j} \leftarrow \{k \in \{1,\ldots,K^{j-1}\}: \|\theta_{y,k}^{\star}\|_{\infty}\leq \delta\}$;  
			\item \textbf{if} $\mathcal{R}^{j}=\emptyset$ \textbf{do}
			\begin{enumerate}[label=\theenumii{}.\arabic*., ref=\theenumii{}.\theenumiii{}]	
				\item \label{step:clustbuilt}$\mathcal{C}_{\emptyset}^{j}\!\leftarrow\!\{k \in \{1,\ldots,K^{j\!-\!1}\}: \#\mathcal{C}_{k} \leq 0.01 \cdot T \}$;
				\item  \label{step:modup} $K^{j} \leftarrow K^{j-1}-K_{\emptyset}^{j}$, with $K_{\emptyset}^{j}=\#\mathcal{C}_{\emptyset}^{j}$;
			\end{enumerate}
			\item \textbf{else}
				\begin{enumerate}[label=\theenumii{}.\arabic*., ref=\theenumii{}.\theenumiii{}]	
				\item\label{step:red} $K^{j} \leftarrow K^{j-1}-K_{R}^{j}$, with $K_{R}^{j}=\#\mathcal{R}^{j}$;
			\end{enumerate}
		\item \textbf{end if};
			\end{enumerate}
		\item \label{step:stop crit1} \textbf{until} $K^{j}=K^{j-1}$
	\end{enumerate}
	\vspace*{.1cm}\hrule\vspace*{.1cm}
	~~\textbf{Output}: Number of sub-models $K^{\star}=K^{j}$.
\end{algorithm}

\subsection{Model order selection via the Elastic Net}
\MM{In this section, we present the algorithm to estimate the model orders  $n_{a}$ and $n_{b}$ of the local affine sub-models, assuming the number of modes $K$ and upper bounds on the parameters $n_{a}$ and $n_{b}$ are known.}

\MM{To this end, the regularization term $r_{y}(\Theta_{y})$ needs to be selected in order \emph{not} to shrink the entire parameter vector to zero (as done in the previous subsection), but only its \emph{redundant} components.} Accordingly, we exploit \emph{elastic net} regularization by setting $r_{y}(\Theta_{y})$ as:
\begin{equation}\label{eq:l_2_1}
r_{y}(\Theta_{y})=\mu \sum_{k=1}^{K} \|\theta_{y,k}\|_{2}^{2}+\nu \sum_{k=1}^{K} \|\theta_{y,k}\|_{1},
\end{equation} 
with $\mu, \nu \in \mathbb{R}^{+}$ being hyper-parameters to be tuned to balance the effect of the two regularization terms. This choice allows us to still enforce parameter selection, thanks to the shrinkage property of the \emph{group lasso} term which shrinks to zero all the parameters associated to unnecessary terms in the regressor, while the presence of the quadratic penalty leads to a more stable and efficient regularization path \cite{Zou2005}.

As the regularization strategy might not lead to parameters that are exactly identical to zero, we remove all terms in the regressor but the one associated to the affine term, for which
\begin{equation}\label{eq:check_2}
\|[\theta_{y,k}]_{i}\|\leq \delta, \quad i=1,\ldots,n_{a}+n_{b}
\end{equation}
where $[\theta_{y,k}]_{i}$ denotes the $i$-th component of $\theta_{y,k}$ and $\delta \in \mathbb{R}^{+}$ is a hyper-parameter to be tuned, accounting for the fact that the regularization strategy might not lead to parameters that are exactly identical to zero. With this we imply that the affine term should never be neglected. 

Note that, by testing the condition in \eqref{eq:check_2} independently for all local models, we do not consider that they share the same structure and thus we cannot enforce this property on the PWA map in \eqref{eq:approx_map}. Nonetheless, a possible heuristic to enforce this property on the PWA map \eqref{eq:approx_map} is to select $n_{a}$ and $n_{b}$ as follows:
\begin{subequations}\label{eq:select_criteria}
	\begin{align}
	&\hat{n}_{a} = \max_{k=1,\ldots,K} \#\{i \in \{1,\ldots,n_{a}\}: \|[\theta_{y,k}]_{i}\|\geq \delta\},\\
	&\hat{n}_{b} \!=\!\! \max_{k=1,\ldots,K} \#\{i \in\! \{1,\ldots,n_{b}\}: \|[\theta_{y,k}]_{n_a+i}\|\geq \delta\},
	\end{align} 	
\end{subequations}
\MM{which evaluates condition in \eqref{eq:check_2} for all local models and fixes the model order of all local models to that of the sub-model with the \emph{least} shrinkage.} We stress that this approach might be \emph{conservative}, since model structure selection relies on the characteristics of at most $2$ sub-models. 

{Algorithm~\ref{algo3} summarizes the proposed strategy for estimating the model orders of the local sub-models. Similar to the Algorithm \ref{algo2}}, this strategy relies on recurrent iterations of Algorithm~\ref{algo1} run with $r_{y}(\Theta_{y})$ defined as in \eqref{eq:l_2_1}. We start from an over-parameterized model with parameters $n_{a,max}$ and $n_{b,max}$ and run Algorithm~\ref{algo3} till the model order is invariant within two consecutive iterations. The model order is then updated at steps~\ref{step:selna} and \ref{step:selnb}, according to the conditions given in \eqref{eq:select_criteria}. Note that, since the model order is iteratively refined, at the beginning of every new run the dataset has to be updated along with the model structure, by redefining $\{x_{t}\}_{t=1}^{T}$ according to the definition in \eqref{eq:regressor} (see step~\ref{step:reg_constr} of Algorithm~\ref{algo3}).  

Once the model structure has been retrieved, Algorithm~\ref{algo1} is run once more for $n_{a}^{\star}$ and $n_{b}^{\star}$ to obtain the PWARX model, imposing $\nu=0$ in \eqref{eq:l_2_1}.

\begin{algorithm}[!tb]
	\caption{[Selection of the sub-models structure]}
	\label{algo3}
	~~\textbf{Input}: Dataset $\{Y,U\}$ of length $T$; initial mode sequence $\mathcal{S}^{0}$; $\rho, \mu, \nu, \lambda, \delta \in \mathbb{R}^{+}$; maximum model complexity $n_{a}^{0}=n_{a,max}$ and $n_{b}^{0}=n_{b,max}$.
	\vspace*{.1cm}\hrule\vspace*{.1cm}
	\begin{enumerate}[label=\arabic*., ref=\theenumi{}]
		\item \textbf{for} $j=1,\ldots$ \textbf{do} \vspace{.1cm}
		\begin{enumerate}[label=\theenumi{}.\arabic*., ref=\theenumi{}.\theenumii{}]	
			\item \label{step:reg_constr}\textbf{construct} $X=\{x_{t}\}_{t=1}^{T}$ for $n_{a}\!=\!n_{a}^{j\!-\!1}$ and $n_{b}\!=\!n_{b}^{j\!-\!1}$;
			\item \label{step:algo1ord} \textbf{run} Algorithm~\ref{algo1} for $r_{y}(\Theta_{y})$ in \eqref{eq:l_2_1}; 
			\item \label{step:selna} $n_{a}^{j} \leftarrow \max_{k} \#\{i \in \{1,\ldots,n_{a}^{j-1}\}: \|[\theta_{y,k}^{\star}]_{i}\|\geq \delta\}$;
			\item \label{step:selnb} $n_{b}^{j}\! \leftarrow \!\!\max_{k} \#\{i \in \{1,\ldots,n_{b}^{j\!-\!1}\}\!: \|[\theta_{y,k}^{\star}]_{i+n_{a}^{j\!-\!1}}\|\!\geq \!\delta\};$
			\end{enumerate}
		\item \label{step:stop crit2} \textbf{until} $n_a^{j}=n_a^{j-1}$ \textbf{and} $n_{b}^{j}=n_{b}^{j-1}$
	\end{enumerate}
	\vspace*{.1cm}\hrule\vspace*{.1cm}
	~~\textbf{Output}: Order of the sub-models $n_a^{\star}=n_a^{j}$ and $n_{b}^{\star}=n_{b}^{j}$.
\end{algorithm}

  \MM{
	\begin{remark}
		For a proper choice of the hyper-parameters $\mu, \nu$, the number of iterations required in Algorithms \ref{algo2} and \ref{algo3} might be ideally reduced to $1$. Nevertheless, the iterative nature of Algorithms \ref{algo2} and \ref{algo3} is expected to result in a good selection of the model structure, even when the hyper-parameters are not tuned to optimality. In this way, the burden due to cross-validation/grid search for tuning the hyper-parameters can be reduced. \hfill $\blacksquare$
\end{remark}}
   
\section{Case studies}\label{Sec:4}
In this section, the performance of the proposed model selection strategies is assessed on two benchmark examples, taken from \cite{Bemporad2005}. The first academic example involves the identification of the PWARX system considered in \cite{Bemporad2005}, while the second case study involves modeling the dynamics of a placement process in a pick-and-place machine using experimental data \cite{Juloski2005}. In both case studies, the model structure selection and identification procedures are performed starting from over-parametrized models with $K_{max}=10$ or $n_{a,max}=n_{b,max}=10$, respectively. The tuning parameters $\mu$ and $\nu$ in the regularization terms \eqref{eq:l_infty_2} and \eqref{eq:l_2_1} are selected so that $\mu \in (0,1)$ and
\begin{equation}\label{eq:convex_comb}
\nu=1-\mu,
\end{equation}
respectively. In this way, we reduce the number of tunable parameters, by considering a convex combination of the $L_2$ and lasso regularizations. 

All computations are carried out an a $i7$ $2.8$~GHz Intel core processor running MATLAB 2019a. The local regression problem in Algorithm~\ref{algo1} are solved by using the CVX package \cite{cvx}, while the PWA separator is estimated via the computationally efficient batch approach proposed in \cite{BBPAut15}, by exploiting the auxiliary parameters used in the example reported therein. To improve the quality of the solution and reduce its dependence on the initial conditions, Algorithm~\ref{algo1} is always executed for $N=20$ different initial mode sequences $\mathcal{S}^{0}$ and the PWARX model is selected as the one leading to the minimum cost.     

\subsection{Numerical example: identification of a PWARX system}
Consider the following PWARX data-generating system with $K=3$ operating conditions
\begin{equation}\label{eq:ex1}
y_{t}=\begin{cases}
\smallmat{-0.4 & 1 & 1.5} \tilde{x}_{t}+e_{t}^{\mathrm{o}}, \quad \mbox{if } \smallmat{4 & -1 & 10}\tilde{x}_{t} < 0,\\
\smallmat{0.5& -1 & -0.5} \tilde{x}_{t}+e_{t}^{\mathrm{o}}, \quad \mbox{if } \smallmat{4 & -1 & 10}\tilde{x}_{t} \geq 0 ~ \&\\
\quad \quad \quad \quad\quad \quad \quad \quad \quad \quad \quad~~  \smallmat{5 & 1 & -6}\tilde{x}_{t} \leq 0\\
\smallmat{-0.3 & 0.5 & -1.7} \tilde{x}_{t}+e_{t}^{\mathrm{o}}, \quad \mbox{if } \smallmat{5 & 1 & -6}\tilde{x}_{t} >0,
\end{cases}
\end{equation}
where
\begin{equation*}
\tilde{x}_{t}=\begin{bmatrix}
y_{t-1} & u_{t-1} & 1
\end{bmatrix}'.
\end{equation*}
The system is excited with a sequence of random \emph{i.i.d.} inputs uniformly distributed in the interval $[-4, \ 4]$ and the corresponding outputs are corrupted by a zero-mean additive noise sequence $e_{o}$, uniformly distributed in the interval $[-0.8, \ 0.8]$. \MM{To asses the performance of the proposed structure selection strategies over different realizations of the training set, we perform a Monte-Carlo simulation with $80$ runs. At each Monte-Carlo iteration, a new dataset of length $T=2000$ is generated with different realizations of the inputs and the measurement noise sequences.} The effect of the noise $e^{o}$ on the measurement output yields an average \emph{signal-to-noise ratio} (SNR), 
\begin{equation}\label{eq:SNR}
\mbox{SNR}=10 \log{\frac{\sum_{t=1}^{T}(y_{t}-e_{t}^{\mathrm{o}})^2}{\sum_{t=1}^{T} (e_{t}^{\mathrm{o}})^2}} = 16~\mbox{dB}.
\end{equation}
\MM{The hyper-parameters are set  to $\lambda=10^{-3}$, $\rho=1$ and $\mu=0.1$ and $\nu$ is selected according to \eqref{eq:convex_comb}.}

\subsubsection{Unknown $K$}
\begin{table*}
	\centering
	\caption{Sub-models parameters (unknown $K$):
		True vs estimated (mean$\pm$standard deviation) over $78$ Monte Carlo runs.}\label{Tab:param1_1}
	\begin{tabular}{|c|c|c|c|c|c|c|}
	\cline{2-7}
	\multicolumn{1}{c}{} & \multicolumn{6}{|c|}{modes}\\
	\cline{2-7}
	\multicolumn{1}{c|}{} & \multicolumn{2}{c|}{$k=1$} & \multicolumn{2}{c|}{$k=2$} & \multicolumn{2}{c|}{$k=3$}\\
	\cline{2-7}
	\multicolumn{1}{c|}{} & true & mean$\pm$std & true & mean$\pm$std & true & mean$\pm$std\\
	\hline
	$[\theta_{y,k}^{\star}]_{1}$ & -0.40 & -0.40$\pm$0.02 & 0.50 & 0.50$\pm$0.01 & -0.30 & -0.30$\pm$0.02\\
	\hline
	$[\theta_{y,k}^{\star}]_{2}$ & 1.00 & 1.00$\pm$0.01 & -1.00 & -1.00$\pm$0.01 & 0.50 & 0.50$\pm$0.01\\
	\hline
	$[\theta_{y,k}^{\star}]_{3}$ & 1.50 & 1.48$\pm$0.07 & -0.50 & -0.50$\pm$0.02 & -1.70 & -1.70$\pm$0.04\\
	\hline
	\end{tabular}
\end{table*}
\begin{table}
	\centering
	\caption{Polyhedral partition parameters (unknown $K$): true vs estimated (mean$\pm$standard deviation) over $78$ Monte Carlo runs.}\label{Tab:param1_2}
	\begin{tabular}{|c|c|c|c|}
		\hline
		\multicolumn{2}{|c|}{first separator} &  \multicolumn{2}{c|}{second separator}\\
		\hline
		true & mean$\pm$std & true & mean$\pm$std\\
		\hline
		4.00 & 4.00$\pm$0.06 & 5.00 & 5.01$\pm$0.08\\
		\hline
		 -1.00 & -1.00$\pm$0.00 & 1.00 & 1.00$\pm$0.00\\
		\hline
		 10.00 & 10.00$\pm$0.13 & -6.00 & -6.01$\pm$0.11\\
		\hline
	\end{tabular}
\end{table}
Let us assume that the \MM{model order of the system} is \MM{known}, namely that $n_{a}=n_{b}=1$, but that the number of local models $K$ is unknown. We iterate Algorithm~\ref{algo2} to estimate the appropriate number of modes, with $\delta$ in \eqref{eq:check1} set to $0.01$. Throughout the Monte Carlo runs only twice the true number of modes $K^{\star}=3$ is not retrieved, in which cases the number of sub-models is estimated to be either $4$ or $5$. 
The average values and standard deviations of the estimated sub-model parameters obtained over the remaining $78$ runs \MM{(for which true number of modes $K^{\star}=3$ has been retrieved)} are reported in \tablename{~\ref{Tab:param1_1}}. These results clearly show that proposed approach allows us to accurately reconstruct the local models. In turn, this implies an accurate reconstruction of the polyhedral partition, as proven by the comparison between the true parameters characterizing the partition \emph{vs} the mean and standard deviation of the ones obtained from the estimated PWA separator in \tablename{~\ref{Tab:param1_2}}. \MM{Overall, the structure of the underlying PWARX system  system has been reconstructed accurately.}  
\subsubsection{Unknown $n_a$ and $n_b$}
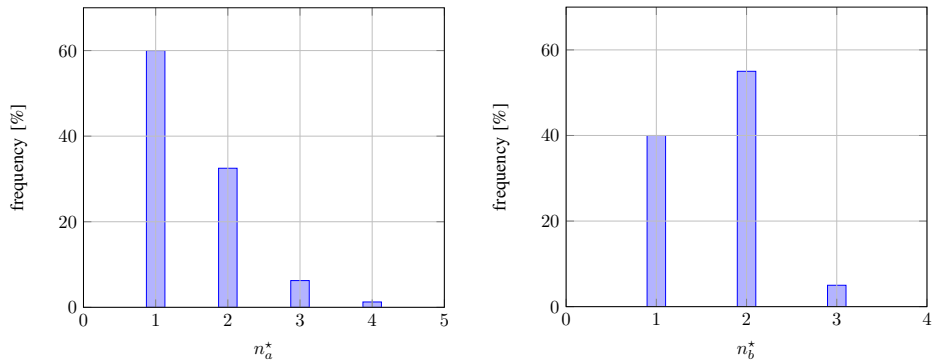
\begin{figure*}
	\centering
	\begin{tabular}{cc}
	\subfigure{\scalebox{0.7}{\begin{tikzpicture}
		\begin{axis}[
		ymin=0, ymax=70,
		area style, 
		xmin=0, xmax=5,
		ylabel={frequency [\%]},
		xlabel={$n_a^{\star}$},
		grid,
		]
		\addplot+[ybar,mark=no] plot coordinates { (1, 60) (2, 32.5) (3, 6.25) (4, 1.25)};
		\end{axis}
		\end{tikzpicture}}} & \subfigure{\scalebox{0.7}{\begin{tikzpicture}
		\begin{axis}[
		ymin=0, ymax=70,
		area style, 
		xmin=0, xmax=4,
		ylabel={frequency [\%]},
		xlabel={$n_b^{\star}$},
		grid,
		]
		\addplot+[ybar,mark=no] plot coordinates { (1, 40) (2, 55) (3, 5)};
		\end{axis}
		\end{tikzpicture}}}
	\end{tabular}
	\caption{Numerical example (fixed number of modes): frequency of the selected orders $n_{a}^{\star}$ and $n_{b}^{\star}$.}\label{Fig:histograms}
\end{figure*}
\begin{table*}
	\centering
	\caption{Sub-models parameters (unknown order):
		True vs estimated (mean$\pm$standard deviation) over $80$ Monte Carlo runs.}\label{Tab:param2_1}
	\begin{tabular}{|c|c|c|c|c|c|c|}
		\cline{2-7}
		\multicolumn{1}{c}{} & \multicolumn{6}{|c|}{modes}\\
		\cline{2-7}
		\multicolumn{1}{c|}{} & \multicolumn{2}{c|}{$k=1$} & \multicolumn{2}{c|}{$k=2$} & \multicolumn{2}{c|}{$k=3$}\\
		\cline{2-7}
		\multicolumn{1}{c|}{} & true & mean$\pm$std & true & mean$\pm$std & true & mean$\pm$std\\
		\hline
		$[\theta_{y,k}^{\star}]_{1}$ & -0.400 & -0.406$\pm$0.022 & 0.500& 0.503$\pm$0.016 & -0.300 & -0.299$\pm$0.014\\
		\hline
		$[\theta_{y,k}^{\star}]_{2}$ & 0.000 & 0.001$\pm$0.007 & 0.000 & 0.001$\pm$0.006 & 0.000 & 0.001$\pm$0.009\\
		\hline
		$[\theta_{y,k}^{\star}]_{2}$ & 0.000 & 0.000$\pm$0.004 & 0.000 & 0.000$\pm$0.002 & 0.000 & 0.001$\pm$0.005\\
		\hline
		$[\theta_{y,k}^{\star}]_{2}$ & 0.000 & 0.000$\pm$0.000 &0.000 & 0.000$\pm$0.000 & 0.000 & 0.000$\pm$0.003\\
		\hline
		$[\theta_{y,k}^{\star}]_{2}$ & 1.000 & 1.000$\pm$0.009 & -1.000 & -1.000$\pm$0.007 & 0.500 & 0.501$\pm$0.006\\
		\hline
		$[\theta_{y,k}^{\star}]_{2}$ & 0.000 & 0.002$\pm$0.011 & 0.000 & -0.001$\pm$0.010 & 0.000 & 0.000$\pm$0.010\\
		\hline
		$[\theta_{y,k}^{\star}]_{2}$ & 0.000 & -0.001$\pm$0.004 & 0.000 & 0.000$\pm$0.003 & 0.000 & 0.000$\pm$0.002\\
		\hline
		$[\theta_{y,k}^{\star}]_{3}$ & 1.500 & 1.475 $\pm$0.079 & -0.500 & -0.497$\pm$0.018 & -1.700 & -1.701$\pm$0.059\\
		\hline
	\end{tabular}
\end{table*}
\begin{table}
	\centering
	\caption{Polyhedral partition parameters (unknown order): true vs estimated (mean$\pm$standard deviation) over $80$ Monte Carlo runs.}\label{Tab:param2_2}
	\begin{tabular}{|c|c|c|c|}
		\hline
		\multicolumn{2}{|c|}{first separator} &  \multicolumn{2}{c|}{second separator}\\
		\hline
		true & mean$\pm$std & true & mean$\pm$std\\
		\hline
		4.000 & 4.001$\pm$0.055 & 5.000 & 5.020$\pm$0.092\\
		\hline
		0.000 & 0.000$\pm$0.010 & 0.000 & 0.000$\pm$0.016\\
		\hline 
		0.000 & -0.001$\pm$0.006 & 0.000 & 0.000$\pm$0.006\\
		\hline 
		0.000 & 0.000$\pm$0.000 & 0.000 & 0.000$\pm$0.000\\
		\hline 
		-1.000 & -1.000$\pm$0.000 & 1.000 & 1.000$\pm$0.000\\
		\hline
		0.000 & -0.002$\pm$0.019 & 0.000 & -0.004$\pm$0.028\\
		\hline 
		0.000 & 0.000$\pm$0.004 & 0.000 & -0.001$\pm$0.005\\
		\hline
		10.000 & 10.000$\pm$0.132 & -6.00 & -6.024$\pm$0.131\\
		\hline 
	\end{tabular}
\end{table}
 \MM{We now assume that the number of sub-models $K=3$ is known, while the model orders $n_{a}, n_{b}$ of the local sub-models are unknown.} Algorithm~\ref{algo3} is iterated over the $80$ Monte Carlo runs to estimate $n_{a}$ and $n_{b}$ along with the parameters of the PWARX model, with $\delta=0.01$. As shown in the histogram reported in \figurename{~\ref{Fig:histograms}}, the \MM{model orders of }  the initial over-parameterized model \MM{(with $n_{a,max}=n_{b,max}=10$) are} shrunk and the final model is  close in structure to the true one. 
 \MM{From \figurename{~\ref{Fig:histograms}}, it can be seen that the proposed approach detected the true model order $n^{\star}_a =1$ and $n^{\star}_b =1$  for about $60\%$ and $40 \%$ of the Monte-Carlo runs, respectively. Note that, for about $58\%$ of the runs, a higher order of the input delays $n_{b} = 2$  is estimated. However, for these cases, as shown in \tablename{~\ref{Tab:param2_1}}, the values of the corresponding model parameters are close to zero, and are negligible with respect to the others.} Therefore, they barely influence the performance of the identified model. 
 We stress that the results shown in \tablename{~\ref{Tab:param2_1}} are obtained after a final iteration of Algorithm~\ref{algo1} with $\nu$ in \eqref{eq:l_2_1} set to zero, and with the model orders selected by running Algorithm~\ref{algo3}.  \MM{The average and standard deviation of the parameters characterizing the regressor space partition are reported  in \tablename{~\ref{Tab:param2_2}}, which are also accurately estimated.} Therefore, despite the slight error in the reconstruction of the model \MM{orders in some of the Monte-Carlo runs, the overall model structure as well as the estimated parameters match closely with that of the true system.}

\subsection{Modeling the dynamics of a pick-and-place process}
\MM{As a second case study, we test the proposed algorithms to model an electronic component placement process in a pick-and-place machine.}
The \MM{set-up} consists of a mounting head carrying an electronic component, that is pushed towards a printed circuit board and released, as soon as the two come in contact \cite{Juloski2005}. The process is characterized by two main operating modes: one associated to the behavior of the pick-and-place machine when it operates in an unconstrained environment and the other when the mounting head comes in contact with the circuit board.

A set of experimental data collected over an interval of $15$~s with a sampling frequency of $400$~Hz is available to model this process (see~\cite{Juloski2005} for further details). This dataset comprises samples of the voltage applied to the motor driving the mounting head of the machine and the vertical position of the mounting head, which represent the input $u$ and output $y$ of the process, respectively. The overall data record is then split into two disjoint sets comprising the $T=4800$ samples associated to the first $12$~s of the experiment and the remaining $\tilde{T}=1200$ samples, that are respectively used to train and validate a PWARX model for the process of interest. 

The regularization parameters used in training are respectively selected as $\rho=2.15 \cdot 10^{-4}$,  and $\mu = 10^{-5}$ (see \cite{BPBB2018}), with $\nu$ chosen according to \eqref{eq:convex_comb}, while $\lambda=10^{-8}$. In validation, the performance attained by simulating the learned PWARX model in open-loop and it is quantitatively assessed via the \emph{best fit rate} (BFR) index, \emph{i.e.,}  
\begin{equation}\label{eq:BFR}
\mbox{BFR}=100 \cdot \max\left\{0,1-\sqrt{\frac{\sum_{t=1}^{\tilde{T}}(y_{t}-\hat{y}_{t})^2}{\sum_{t=1}^{\tilde{T}}(y_{t}-\bar{y})^{2}}}\right\}\%
\end{equation} 
with $\bar{y}$ and $\hat{y}_{t}$ denoting the average output over the validation set and the reconstructed one, respectively.   

\begin{figure*}
	\vspace{-3cm}
	\centering
	\begin{tabular}{cc}
		\subfigure[$K^{\star}=3$ and $n_{a}=n_{b}=2$\label{Fig:PP_performanceK}]{\includegraphics[width=6.3cm,trim=3cm 7cm 3cm 4cm,clip]{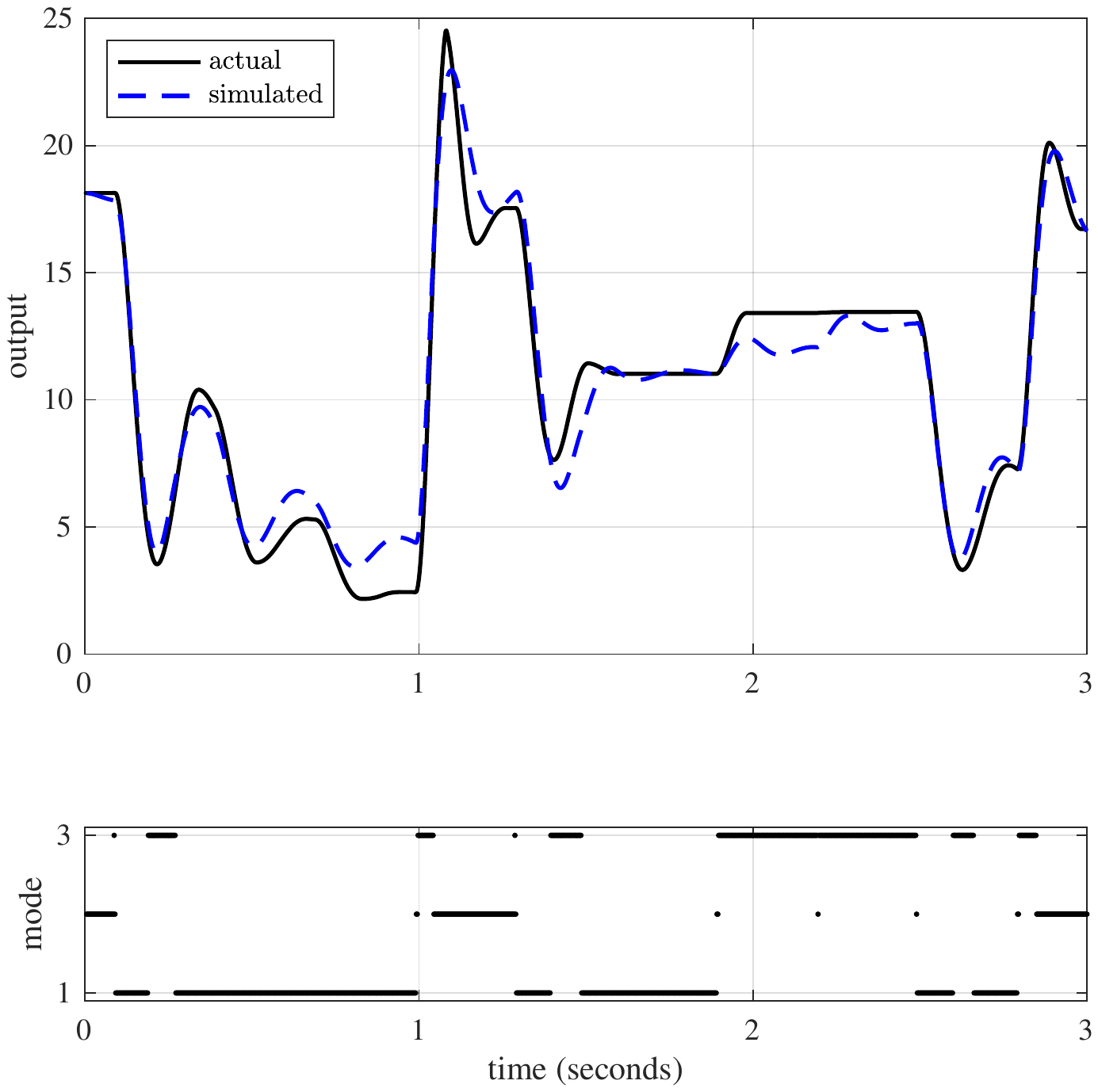}} & \subfigure[$K=2$ and $n_{a}^{\star}=n_{b}^{\star}=2$\label{Fig:PP_performanceor}]{\includegraphics[width=6.3cm,trim=3cm 7cm 3cm 1cm,clip]{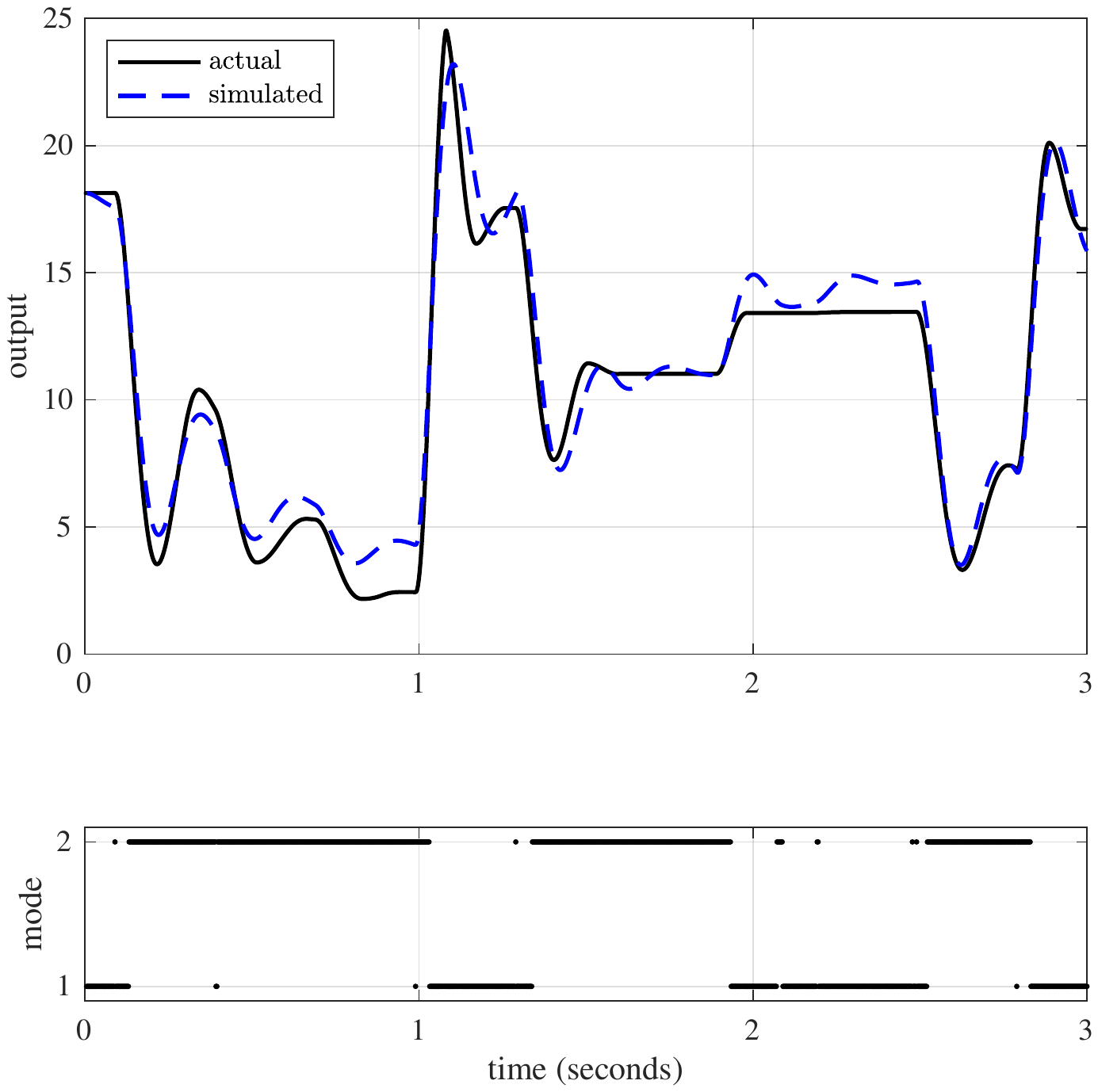}} 
	\end{tabular}
	\caption{Pick-and-place machine: simulated \emph{vs} actual output (top) and reconstructed mode sequence for fixed local model structure (left) and fixed number of sub-models (right).\label{Fig:PP_perofrmance}}
\end{figure*}

\subsubsection{Unknown $K$}
Let $n_{a}$ and $n_{b}$ be fixed according to the order of the PWARX model estimated in \cite{Bemporad2005}, namely $n_{a}=n_{b}=2$. We run Algorithm~\ref{algo2} with $\delta=10^{-3}$ in \eqref{eq:check1}, which terminates after $6$ iterations and it leads to the selection of $K^{\star}=3$ modes. \MM{The proposed structure selection strategy thus introduces an additional mode with respect to the two main operating conditions of the underlying process.} Nonetheless, we drastically reduce the complexity of the model with respect to the initial \MM{over-parameterized model}, characterized by $K_{max}=10$ \MM{modes}. The comparison between the output reconstructed in validation with the learned PWARX model with 3 modes and the actual one is reported in \figurename{~\ref{Fig:PP_performanceK}}, along with the sequence of active local models. This result corresponds to a BFR of $81.2$\%.

\subsubsection{Unknown $n_a$ and $n_b$}
We now assume \MM{the number of local models is fixed to} $K=2$ and we \MM{run} Algorithm~\ref{algo3} to \MM{estimate} the \MM{sub-model orders}. \MM{We set} the threshold for selection in \eqref{eq:check_2} as $\delta=10^{-3}$. Algorithm~\ref{algo3} terminates after $5$ runs and the resulting PWARX model is of the second order, namely $n_{a}^{\star}=n_{b}^{\star}=2$. \MM{Thus, the complexity of the initial over-parameterized model with  $n_{a,max}=n_{b,max}=10$ is shrunk, leading to a much simpler model, which is same as the one used in \cite{Bemporad2005,BPBB2018}.} \figurename{~\ref{Fig:PP_performanceor}} shows the comparison between the simulated and reconstructed output in validation, along with the obtained active mode sequence, which corresponds to a BFR of $79.8$\%. \MM{It can be observed that the estimated PWARX model with simpler model structure is still able to accurately describe the behavior of the  placement process.} 

\section{Conclusions}\label{Sec:Conclusions}
The paper has presented a preliminary study on the application of \emph{regularization-based} strategies to select the model structure of a PWARX model. The proposed approaches combine the selective power of existing shrinkage techniques with heuristics, thus providing a tool for data-driven model structure selection. The results obtained in the considered examples highlight the potential of these strategy. 

Future research will be devoted to find less conservative criteria for the selection of the local model structure and to address a complete structure selection problem. The sensitivity of the presented algorithms w.r.t. a set of hyper-parameters will be analyzed and  auto-tuning procedures for this parameters will be investigated. Thanks to the generality of the \emph{coordinate-descent} approach on which our strategies rely, future work will also include their extension to other classes of switching models.

\bibliographystyle{plain}
\bibliography{PWAidsel}


\end{document}